\documentclass[fleqn,usenatbib]{mnras}
\usepackage{natbib}

\usepackage[T1]{fontenc}
\usepackage{ae,aecompl}

\usepackage{graphicx}	
\usepackage{amsmath}	
\usepackage{amssymb}	
\usepackage{pdflscape}	
\usepackage{float}
\usepackage{scrextend}

\usepackage{txfonts}
\usepackage{footmisc}
\usepackage{color}
\usepackage[exponent-product=\cdot]{siunitx}

\usepackage{multirow}
\usepackage{subcaption}
\usepackage[switch]{lineno}
\usepackage{booktabs}
\usepackage{longtable}
\usepackage[para,flushleft]{threeparttable}

    \makeatletter
    \renewcommand*{\@fnsymbol}[1]{\ensuremath{\ifcase#1\or  \or \dagger\or \ddagger\or
        \mathsection\or \mathparagraph\or \|\or **\or \dagger\dagger
        \or \ddagger\ddagger \else\@ctrerr\fi}}
    \makeatother

\title[M~87 low state 2012-2015]{Monitoring of the radio galaxy M~87 during a low emission state from 2012 to 2015 with MAGIC}

\author[V.~A.~Acciari~et.~al.]{\parbox{\textwidth}{\Large{
MAGIC Collaboration: V.~A.~Acciari$^{1}$,
S.~Ansoldi$^{2,23}$,
L.~A.~Antonelli$^{3}$,
A.~Arbet Engels$^{4}$,
C.~Arcaro$^{29,30\textcolor{blue}{\star}}$,
D.~Baack$^{5}$,
A.~Babi\'c$^{6}$,
B.~Banerjee$^{7}$,
P.~Bangale$^{14\textcolor{blue}{\star}}$,
U.~Barres de Almeida$^{8}$
% ...
}
\newline
\emph{\normalsize Affiliations are listed at the end of the paper}
}}

\date{Accepted XXX. Received YYY; in original form ZZZ}

\pubyear{2019}

\newcommand\blfootnote[1]{%
   \begingroup
   \renewcommand\thefootnote{}\footnote{#1}%
   \addtocounter{footnote}{-1}%
   \endgroup
    }

\begin{document}
\label{firstpage}
\pagerange{\pageref{firstpage}--\pageref{lastpage}}
\maketitle
\clearpage
\begin{abstract}
M\,87 is one of the closest (z=0.00436) extragalactic sources emitting at very-high-energies (VHE, $E>100$\,GeV).
The aim of this work is to locate the region of the VHE gamma-ray emission and to 
describe the observed broadband spectral energy distribution (SED) during the low VHE gamma-ray state.
The data from M\,87 collected between 2012 and 2015 as part of a MAGIC monitoring programme are analysed and combined with multi-wavelength data from \textit{Fermi}-LAT, \textit{Chandra}, HST, EVN, VLBA and the Liverpool Telescope.
The averaged VHE gamma-ray spectrum can be fitted from $\sim$100\,GeV to $\sim$10\,TeV with a simple power law with a photon index of $(-2.41\pm0.07)$, while the integral flux above 300\,GeV is $(1.44\pm0.13)\times10^{-12}\,\mathrm{cm}^{-2}\,\mathrm{s}^{-1}$. During the campaign between 2012 and 2015, M\,87 is generally found in a low emission state at all observed wavelengths. The VHE gamma-ray flux from the present 2012-2015 M\,87 campaign is consistent with a constant flux with some hint of variability ($\sim3\,\sigma$) on a daily timescale in 2013. 
The low-state gamma-ray emission likely originates from the same region as the flare-state emission. Given the broadband SED, both a leptonic synchrotron self Compton and a hybrid photo-hadronic model reproduce the available data well, even if the latter is preferred. We note, however, that the energy stored in the magnetic field in the leptonic scenario is very low suggesting a matter dominated emission region.\newline
\textbf{Keywords:} galaxies: active - galaxies: individual: M~87 - galaxies: jets - gamma rays: galaxies - radiation mechanisms: non-thermal
\end{abstract}

\section{Introduction}\label{sec:intro}
M\,87\blfootnote{\textcolor{blue}{$^{\star}$} Corresponding authors (alphabetical ordered): C.~Arcaro, P.~Bangale, M.~Manganaro, D.~Mazin.~\href{mailto:contact.magic@mpp.mpg.de}{contact.magic@mpp.mpg.de}
\newline \textcolor{blue}{$^{\star\star}$}: deceased} is a large elliptical radio galaxy of Fanaroff-Riley type I ~\citep[FR I;][]{1974MNRAS.167P..31F}, located in the Virgo Cluster, at a distance of $(16.4\pm0.5)$\,Mpc \citep{2010A&A...524A..71B}.
M\,87 is powered by a super-massive black hole (SMBH) with a mass assumed to be $(6.5\pm0.2_{\mathrm{stat}}\pm0.7_{\mathrm{sys}}) \times 10^{9} \mathrm{M}_{\odot}$ \citep{2011ApJ...729..119G,2019ApJ...875L...6E}.\par
The relativistic jet of M\,87 is misaligned with respect to our line of sight with an angle between 15-25$^{\circ}$ \citep{1999ApJ...520..621B,2009Sci...325..444A,2018ApJ...855..128W}.
The orientation and vicinity of M\,87 allow the jet to be studied during its evolution, from the core to the extended lobe, where it slows down ending its path interacting with the intergalactic medium~\citep{2000ApJ...543..611O}.
The jet formation of M\,87 has been recently studied by \citep{2018A&A...616A.188K} in the mm range at 86\,GHz, revealing  a parabolically expanding limb-brightened jet which emanates from a resolved VLBI core of  $\sim (8-13)$ Schwarzschild radii in size.
In \citep{2018A&A...610L...5K} a strong indication was found that the jet base does not consist of a simple homogeneous plasma, but of inhomogeneous multi-energy components.
The jet extends for $\gtrsim 30\arcsec$ \citep{2002ApJ...564..683M}, and several knots along its length have been resolved in radio, optical and X-ray bands \citep{2001ApJ...551..206P,2002ApJ...568..133W}. The inner knot HST-1, located at $0\,\farcs85$ from the core, has been in a flaring state since 2000 \citep{2003ApJ...586L..41H,2005ApJ...624..656W,2006ApJ...640..211H}, reaching the maximum flux in 2005 and had a secondary flaring in 2006-2007 \citep[see][]{2009ApJ...699..305H,2009AJ....137.3864M,2012ApJ...746..151A}.\par
The temporal correlation between the very-high-energies (VHE, E\,$>$\,100~GeV) gamma-ray emission and multi-wavelength (MWL) data, in which the source is spatially resolved, provides a unique opportunity to locate the origin of VHE gamma-ray emission in active galactic nuclei (AGNs). M87 was detected in TeV gamma rays first by the HEGRA (High-Energy-Gamma-Ray Astronomy) Collaboration in 1998~\citep{2003A&A...403L...1A}. The VERITAS (Very Energetic Radiation Imaging Telescope Array System) Collaboration reported a clear detection of M\,87 in the 2007 campaign at energies above 250\,GeV~\citep{2008ApJ...679..397A} and continued to routinely monitor the source~\citep{2010ApJ...716..819A}. The first detection of gamma-ray emission from M\,87 with MAGIC (Major Atmospheric Gamma-ray Imaging Cherenkov)  occurred in 2005 in a low flux state, and results of those observations together with those performed between 2006 and 2007 were reported in~\citet{2012A&A...544A.142A}. \par
During a flare in 2008, detected through a monitoring campaign, MAGIC observed a flux variability on timescales as short as a day~\citep{2008ApJ...679..397A,2008ApJ...685L..23A}.
As of 2019, M\,87 has been monitored for more than 10 years in the TeV band by MAGIC, H.E.S.S. (High Energy Stereoscopic System), and VERITAS~\citep{2009Sci...325..444A, 2012ApJ...746..151A,2012ApJ...746..141A,2012AIPC.1505..586B}. According to the available VHE gamma-ray data, a total of three periods of high activity occurred in 2005, 2008 and 2010. \par
The modelling of VHE gamma-ray emission in the context of the broadband spectral energy distribution (SED) is challenging and draws the attention of several theory groups, see, e.g., \citet{2005ApJ...634L..33G,2008MNRAS.385L..98T,2005A&A...432..401G, 2009Ap&SS.321...57I,2010MNRAS.402.1649G}. \par
The low state of M\,87 is important for modelling as it can be used to describe a "baseline" state to be used as a reference, even if several model parameters remain unconstrained in the absence of flux variability. The study of the source in a high state can then be associated to the low-state reference improving the interpretation of the emission scenarios. ~\citet{2012A&A...544A.142A} previously modeled the low-state broadband SED using the same model applied to the high states observed in 2008. \cite{2015MNRAS.450.4333D} and \citet{2016MNRAS.457.3801P} also studied the M\,87 activity in a MWL context. \citet{2016MNRAS.457.3801P} singled out two different states of the source, a low and a more active one and studied them separately.\par
In this paper, the MAGIC monitoring dataset of M\,87 between 2012 and 2015 is presented. No major flare is detected in this period, which allows us to study the source in low flux state. The data quality is sufficient to constrain some emission models and study the MWL SED of M\,87 from radio to  VHE gamma-ray frequencies, using MAGIC and available MWL data.\par
The paper is structured as follows: The observations and data analysis for the several instruments involved are presented in section~\ref{sec:obs}. The results, consisting of the long-term light curves, skymaps and SEDs of the source in a MWL context are reported and described in detail in section~\ref{sec:res}. The SED modelling is discussed in section~\ref{sec:sed} and conclusions are summarized in section~\ref{sec:conclusions}. 

\section{Observations \& analysis} \label{sec:obs}
In the following, the data collected within this MWL campaign, ordered from the highest (gamma rays; MAGIC) to lowest energies (radio; VLBA --Very Long Baseline Array), are presented.
\subsection{MAGIC} \label{sec:magic} 
MAGIC is a stereoscopic system of two 17-m diameter imaging atmospheric Cherenkov telescopes situated at the Roque de los Muchachos, on the Canary island of La Palma. An integral sensitivity corresponding to $(0.66 \pm 0.03)$\% of the Crab Nebula flux above 220\,GeV  is achieved in 50\,hrs at low zenith angles \citep[see][for details on the telescopes performance]{2016APh....72...76A}.\par
M\,87 observations were performed regularly during the visibility period between December and July in the years 2012 to 2015. The observations took place at zenith angles ranging from $15^{\circ}-50^{\circ}$ during dark time and under moonlight conditions.
Data were analysed using the standard MAGIC reconstruction software \citep[MARS;][]{zanin2013}.
Further details on the stereo MAGIC analysis and on the telescopes performance under moonlight can be found in~\citet{2016APh....72...76A} and \citet{2017APh....94...29A}, respectively.

\subsection{{\it Fermi}-LAT}\label{sec:fermi}
The Large Area Telescope (LAT) on board the {\it Fermi} satellite is a pair-conversion telescope that covers the energy range from 20\,MeV to more than 300\,GeV, with an angular resolution of $\theta_{68\%} = 0.8^{\circ}$ at 1\,GeV and a field of view of 2.4\,sr~\citep{2009ApJ...697.1071A}. 
The unbinned likelihood analysis of the {\it Fermi}-LAT data was based on the publicly available Pass~8 photon dataset\footnote{\href{http://fermi.gsfc.nasa.gov/cgi-bin/ssc/LAT/LATDataQuery.cgi}{http://fermi.gsfc.nasa.gov/cgi-bin/ssc/LAT/LATDataQuery.cgi}}. The data were analysed with the \textit{Fermi} Science Tools package (version v10r0p5), using the \texttt{Source} (P8R2\_SOURCE\_V6) event class. The M\,87 light curve was constructed for $\mathrm{E} > 300$ MeV with 30-days time bins. All events within $8^\circ$ of the region of interest centered at the catalog position of M\,87 were selected. A dedicated likelihood analysis was performed on each time bin. All point sources from the Large Area Telescope source catalog \citep[3FGL;][]{2015ApJS..218...23A} that lied within a $12^\circ$ circle from M\,87 were included in the model over each time interval. 
The resulting average flux was found to be $(6.85 \pm 0.56) \times 10^{-9}~\mathrm{cm}^{-2}~\mathrm{s}^{-1}$.
{\it Fermi}-LAT flux densities and the energy spectrum were found both to be consistent with the values reported in the 3FGL.

\subsection{{\it Chandra}}\label{sec:chandra} 
The data reduction procedure of the X-ray dataset was performed following the {\it Chandra} Interactive Analysis of Observations (CIAO) threads\footnote{\href{http://cxc.harvard.edu/ciao/guides/index.html}{http://cxc.harvard.edu/ciao/guides/index.html}}, using CIAO version 4.7 and {\it Chandra} Calibration Database (CALDB) version 4.6.9. 
The X-ray images were `registered' aligning the nuclear X-ray position to the location of the radio core (see for additional details \citealt{2010ApJ...714..589M,2012ApJS..203...31M,2013ApJS..206....7M}).\par
To measure observed fluxes for the nuclear emission as well as for any jet feature, a region of size and shape appropriate to the observed X-ray emission was chosen. The background contamination was estimated using two regions with the same shape and size as the science targets but offset from the jet~\citep{2015ApJS..220....5M}. \par
A 1\,$\sigma$ error was calculated based on the square root of the number of counts (the standard deviation of a Poissonian distribution) in the source and background regions. Fluxes reported here were also corrected for the Galactic absorption assuming a photon index of -2 and a value for the Galactic column density of $1.94\times10^{-20}$\,cm~\citep{2005A&A...440..775K}.\par
More details on the X-ray data reduction and analysis can be found in \cite{2009ApJ...692L.123M,2009ApJ...696..980M,2011ApJS..197...24M}.

\subsection{HST}\label{sec:hst}
HST (Hubble Space Telescope) data presented here were obtained with the Space Telescope Imaging Spectrograph (STIS). These imaging data were obtained using the Near-Ultraviolet (NUV) Multi-Anode Microchannel Array (MAMA) detector of STIS, which has a pixel scale of 0\,{\farcs}024 per pixel providing the best resolution currently available with HST~\citep{2019AAS...23315706M}. The filter in use for these observations was the F25QTZ filter which is a long-pass quartz filter centered at 2360\,$\AA$ with a FWHM of about 1000\,$\AA$. More details on the sensitivity and throughput of this filter are given in the STIS instrument handbook~\citep{Hernandez2014}.\par
Aperture corrections were applied following~\cite{2003stis.rept....1P}. In addition, an extinction correction of A(F25QTZ)=0.190\,mag~\citep{1989ApJ...345..245C} was applied and fluxes were converted to erg\,cm$^{-2}$\,s$^{-1}$\,$\AA^{-1}$.\par 
The HST-1 knot located 0\,{\farcs}85 from the center of the galaxy was clearly distinguished from the AGN, as well as other knots ``downstream''. A more detailed account of recent observations of M\,87 with the HST is given in~\citet{2009AJ....137.3864M} and~\citet{2011ApJ...743..119P}. 

\subsection{Liverpool Telescope}\label{sec:pol}
The optical polarization data were taken with the 2-m Liverpool Telescope \citep[LT;\,][]{2004SPIE.5489..679S} located on the Canary island of La Palma. The 2012 observations were performed as a part of the Ringo2 blazar programme ~\citep{2016MNRAS.462.4267J} and 2014-2015 observations using the Ringo3 polarimeter \citep{2012SPIE.8446E..2JA}. Ringo2 observations were performed using a V+R hybrid filter. \par
The Ringo3 polarimeter consists of a rotating polaroid (1 rotation every 4 seconds) which captures 8 images of the source at successive 45$^\circ$ rotations of the polaroid. These 8 exposures could be combined according to the equations in ~\citet{2002A&A...383..360C} to determine the degree and angle of polarization. Ringo3 data are separated into three wavelengths bands using dichroic mirrors rather than standard filters.\par
A gap between Ringo2 and Ringo3 data in the present work was due to the time needed for the upgrade of the polarimeter.

\subsection{EVN}\label{sec:env}
The {\bf E}uropean {\bf V}ery Long Baseline Interferometer (VLBI) {\bf N}etwork (EVN\footnote{\href{http://www.evlbi.org}{http://www.evlbi.org}}) is an interferometric array of radio telescopes spread throughout Europe (and beyond). It conducts unique high-resolution radio astronomical observations of cosmic radio sources. \par
Radio flux densities were taken with EVN during 2012-2015. M\,87 was observed with EVN for a total of 10 epochs between 2012 January and 2015 May as a part of the long-term M\,87/HST-1 monitoring project starting from mid-2009 \citep{2011IAUS..275..150G,2012A&A...538L..10G,2014ApJ...788..165H,2015arXiv150401808H}. For all the observations, the data were recorded  and correlated at the Joint Institute for VLBI in Europe (JIVE; see \citealt{2012A&A...538L..10G,2014ApJ...788..165H} and \citealt{2015arXiv150401808H} for some more detailed information). The initial data calibration and fringe-fitting was performed in Astronomical Image Processing System (AIPS\footnote{\href{http://www.aips.nrao.edu/index.shtml}{http://www.aips.nrao.edu/index.shtml}}) based on the standard VLBI reduction procedures. The final images were produced in the DIFMAP software \citep{1994BAAS...26..987S} after several cycles of phase and amplitude self-calibration. For the core, the peak flux densities were convolved with a 5/10-milliarcseconds (mas) as a circular Gaussian beam for 5-GHz/1.7-GHz data, respectively.

\subsection{VERA}\label{sec:vera}
Between September 2011 and September 2012, the core of M\,87 was densely monitored with the VLBI Exploration of Radio Astrometry (VERA), a Japanese VLBI network consisting of 4 stations operated at 22 and 43\,GHz. A total of 24 epochs were obtained at 22\,GHz throughout the period, and additional 5 sessions were performed at 43\,GHz between February and May 2012. Detailed descriptions of the data analysis as well as some images were presented in \citet{2014ApJ...788..165H}. In the present paper, an improved amplitude calibration procedure was applied to better take into account the amplitude loss due to multiple signal digitization processes during data recording \citep{2005PASJ...57..259I}. Peak flux densities of the M87 core (at 22 and 43\,GHz) were provided, that were measured with a common 0.6-mas-diameter circular Gaussian convolving beam. An amplitude uncertainty of 10\% for each dataset was assumed. 

\subsection{VLBA}\label{sec:vlba}
During the period discussed in this paper, five high-resolution radio observations of M\,87 were made at 43\,GHz using the VLBA~\citep{1994IEEEP..82..658N}. Those observations were designed to monitor the ambient structure of M\,87 in support of observations that would have been made in response to a flare in the gamma-ray energy band.  No such flare occurred. A major upgrade to the VLBA digitization and recording hardware was occurring during this period\footnote{Dominici Science Operations Center 2014, VLBA Observational Status Summary 2015A (Socorro, NM:NRAO);\href{https://science.lbo.us/facilities/vlba/docs/manuals/oss2015A}{https://science.lbo.us/facilities/vlba/docs/manuals/oss2015A}} which provided an improvement in sensitivity by a factor of two. Changes also occurred in the flux-density calibration methods and standards used~\citep{Walker2014}. The data were processed in AIPS using standard methods. 
Additional details about the reduction of the VLBA data, along with the imaging and analysis results, could be found in \citet{2018ApJ...855..128W}.

\section{Results}\label{sec:res}
In the following the results of this MWL campaign are presented, starting with the detection in the TeV band by MAGIC in Section ~\ref{subsec:magicresults}, followed by the discussion on the MWL light curves in Section~\ref{subsec:lightcurves}, and concluding with the characterization of the SED in the GeV-TeV band in Section~\ref{subsec:spectrum}.

\subsection{Detection and sky maps with MAGIC} \label{subsec:magicresults}
MAGIC has detected M\,87 in every yearly campaign between 2012 and 2015. Table~\ref{sigma-table} lists the effective observation time and significance of the VHE gamma-ray signal. The significance of the detection is calculated according to Eq.~17 in~\citet{1983ApJ...272..317L}.
\begin{table}
\caption{Effective observation time and significance of the VHE gamma-ray signal observed from M\,87 above 300\,GeV between 2012 and 2015.}\label{sigma-table}
\begin{center}
\begin{tabular}{c|c|c}
\hline\hline
Year&T$_{\mathrm{eff}}$ [hrs]& Significance$_{E>300\,\mathrm{GeV}}$ [$\sigma$]\\
\hline
\hline
2012 & 38.75 & 5.41\\
2013 & 34.82 & 8.75\\
2014 & 49.88 & 7.29\\
2015 & 32.72 & 5.96\\
\hline
\hline
\end{tabular}
\end{center}
\end{table}
  \begin{figure}
  \centering
 \resizebox{\hsize}{!}{\includegraphics{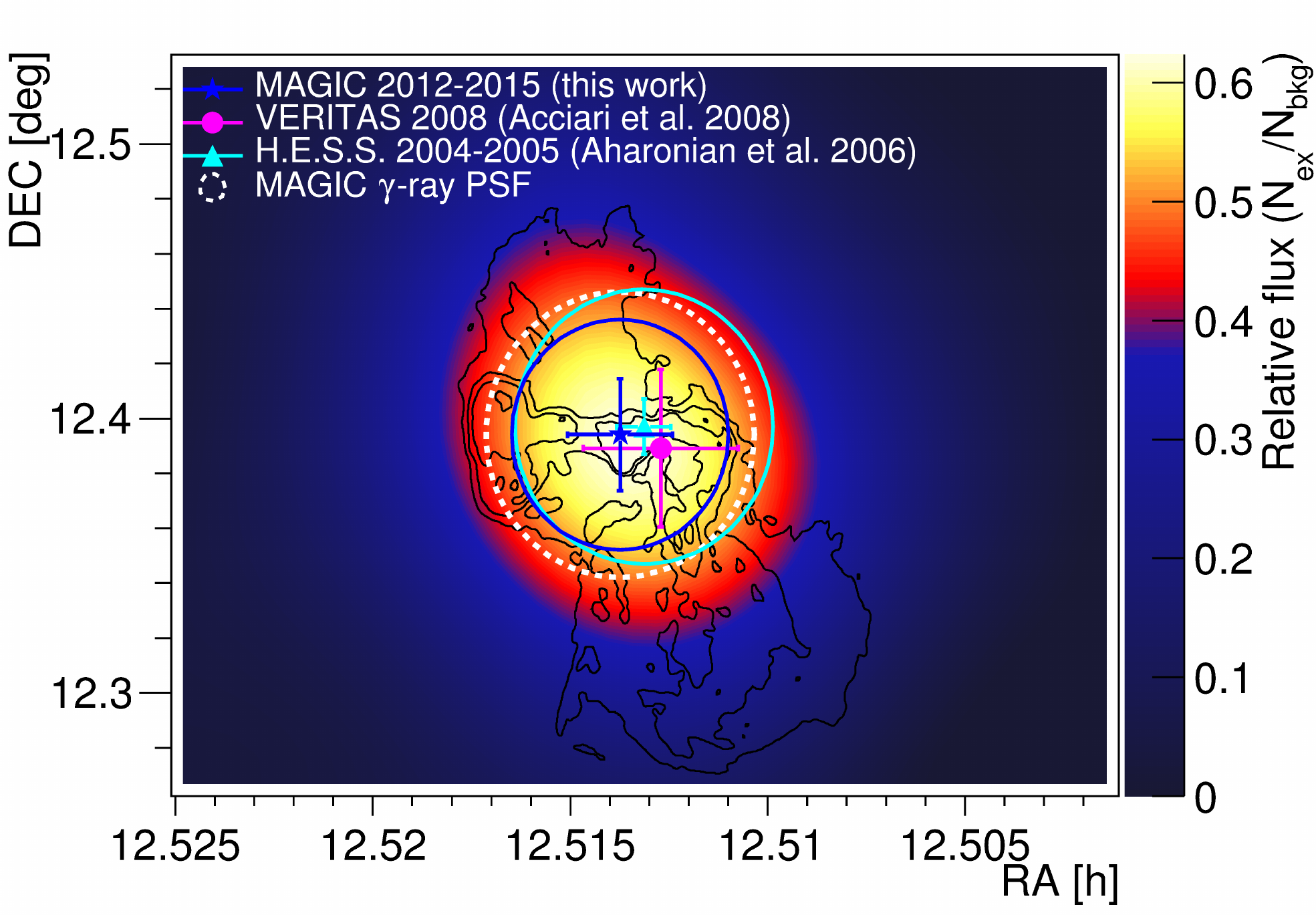} }
   \caption{Averaged VHE gamma-ray sky map above 300\,GeV of M\,87 derived from 2012 to 2015 MAGIC stereo observations centered on the position of the VHE gamma-ray emission obtained from a two-dimensional Gaussian fit (blue star). The PSF (white dashed circle) of the MAGIC telescopes for a gamma-ray signal is shown. The positions fitted to the VHE gamma-ray signal observed by MAGIC from 2012 to 2015 (blue star), by H.E.S.S. from 2004 to 2005 (light blue triangle;~\citealt{2006Sci...314.1424A}) and by VERITAS in 2007 (magenta point;~\citealt{2008ApJ...679..397A}) are shown. The circles (solid blue and light blue lines) indicate the 99.9\% confidence-level upper limit of an extended gamma-ray signal from MAGIC and H.E.S.S. observations, respectively.
   The VLA radio image (black contours) at 327\,MHz~\citep{2000ApJ...543..611O} is shown as a reference.}
   \label{fig:VHEskymap} 
 \end{figure}
In Fig.~\ref{fig:VHEskymap} the contours of the radio image taken with VLA (Very Large Array) at 327\,MHz~\citep{2000ApJ...543..611O} are superimposed on the VHE gamma-ray relative-flux sky map, showing the corresponding PSF of the MAGIC telescopes (0.052$^\circ$) and the upper limit at 99.9\% confidence level of an extended VHE gamma-ray signal which is of the size of $0.042^{\circ}$, corresponding to 11.5\,kpc. An enhanced point-like gamma-ray signal is found at the position of the catalog position of M\,87. 
The locations of the VHE gamma-ray emission observed by H.E.S.S.~\citep{2006Sci...314.1424A} and VERITAS~\citep{2008ApJ...679..397A} are indicated in Fig.~\ref{fig:VHEskymap} as well.
 \begin{table}
 \caption{Mean integral flux above 300\,GeV observed with MAGIC between 2012 and 2015 obtained from a fit with a constant to the $\sim$20-day binned light curves.}\label{lc1-table}
\begin{center}
\begin{tabular}{c|c}
\hline
\hline
Year & $F_{>300\,\mathrm{GeV}}$ [$10^{-12}$ cm$^{-2}$ s$^{-1}$] \\
\hline
\hline
2012 & $1.18\pm0.25$ \\ 
2013 & $1.72\pm0.30$ \\ 
2014 & $1.49\pm0.22$ \\ 
2015 & $1.25\pm0.33$ \\ 
2012-2015 & $1.44\pm0.13$ \\ 
\hline
\hline
\end{tabular}
\end{center}
\end{table}
\begin{figure*}
  \centering
     \includegraphics[width=17cm]{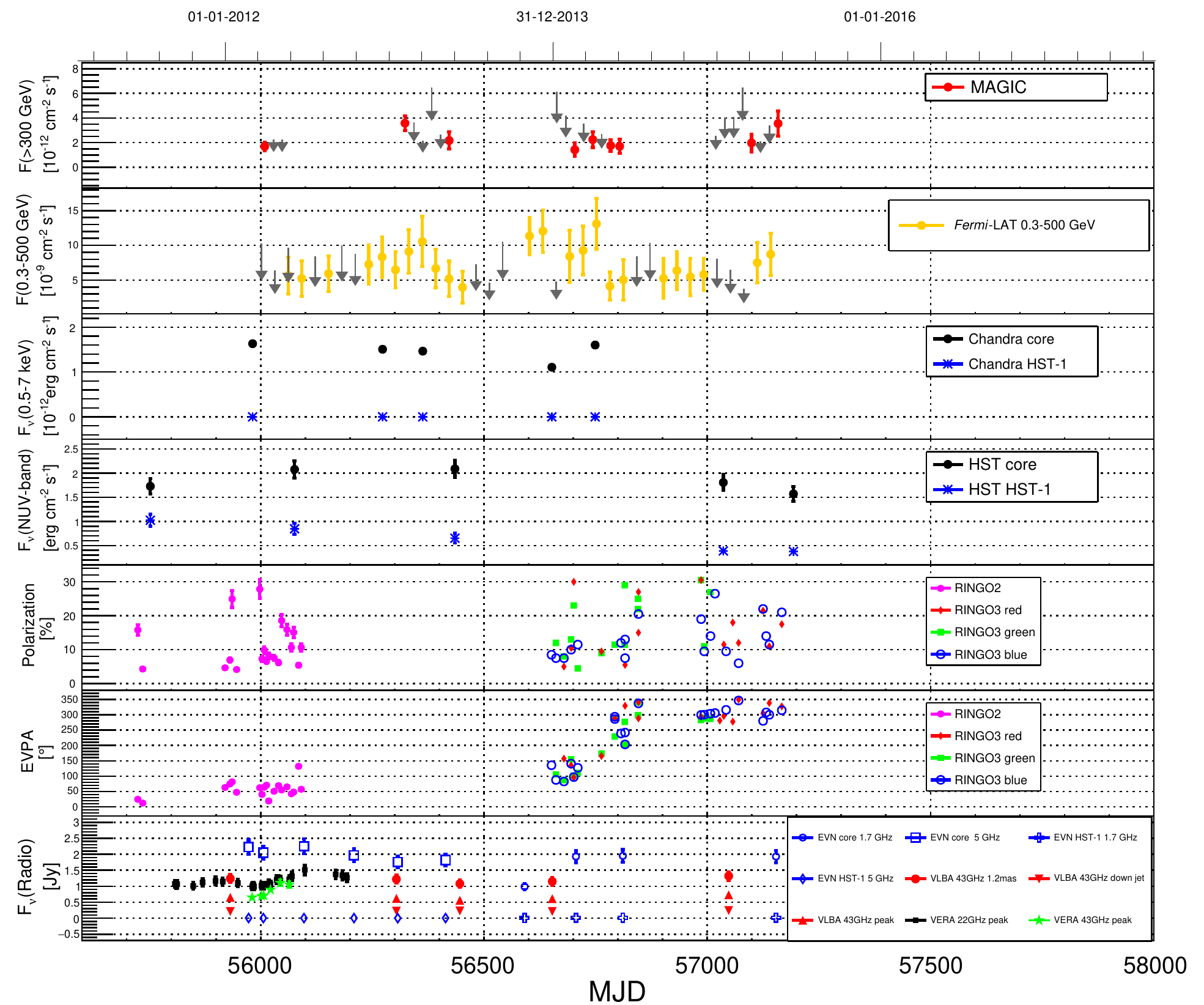}
    \caption{MWL light curve between 2012 and 2015: \textit{From top to bottom:} VHE and HE gamma-ray data by MAGIC and \textit{Fermi}-LAT, respectively. Upper limits at the 95\% confidence level are indicated by downward gray arrows (see text for details). The \textit{Chandra} X-ray fluxes from the core (points) and HST-1 (asterisks) are shown in the third panel from top. The X-ray observations are corrected for Galactic absorption (see text for details). In the fourth panel from top the NUV data from the core as full points and from HST-1 as stars are presented. NUV data have been corrected for interstellar extinction following~\citet{1989ApJ...345..245C}. Optical polarization data taken with V+R filter by the Liverpool telescope with RINGO2 in 2012 (filled circles) and with RINGO3 in 2014-2015 (empty circles, filled squares and diamonds for blue, green and red band respectively) are shown in the dedicated panels. The bottom panel presents radio data provided by the EVN (blue open symbols), VLBA (red filled symbols), and VERA (black empty and filled stars). The light curves are daily binned except VHE and HE gamma rays, where roughly 20- and 30-days binning is applied, respectively.}
    \label{mwl_lc}
   \end{figure*}
\begin{table*}
\caption{Comparison of the integral fluxes ($400\,\mathrm{GeV}<E<1$\,TeV), spectral indices and differential fluxes observed in 2004-2005 with H.E.S.S.~\citep{2006Sci...314.1424A}, 2005-2007 with MAGIC~\citep{2012A&A...544A.142A}, in 2007 and 2010
with VERITAS~\citep{2008ApJ...679..397A,2012ApJ...746..141A}, in 2008 with MAGIC~\citep{2008ApJ...685L..23A} and between 2012 and 2015 with MAGIC (this work). The integral fluxes are extrapolated from the simple power-law fits to the observed spectra.  The differential flux is compared for the decorrelation energy of 784\,GeV obtained for the data of this work, adopting the parameters of the individual fits. The values shown include statistical errors only and are obtained by error propagation assuming the errors of the flux normalization and the spectral index to be uncorrelated.}
\label{tab:spec}
\begin{center}
\begin{tabular}{c|c|c|c|c}
\hline\hline
Array & Year & $F_{400\,\mathrm{GeV}<E<1\,\mathrm{TeV}}$  [$10^{-12}$\, cm$^{-2}$\,s$^{-1}$]
&$\Gamma$ & $f_{E=784\,\mathrm{GeV}}$
 [$10^{-12}$ TeV$^{-1}$ cm$^{-2}$ s$^{-1}$] \\
\hline
\hline
H.E.S.S. & 2004    & $0.51\pm0.22$  & $-2.62\pm0.35$  & $0.46\pm0.15$ \\
\hline
H.E.S.S. & 2005  & $1.97\pm0.44$ & $-2.22\pm0.15$  & $2.01\pm0.28$ \\
\hline
MAGIC & $2005-2007$  & $0.90\pm 0.44$   & $-2.21\pm0.21$  & $0.92\pm0.24$\\
\hline
VERITAS & 2007 & $1.31\pm0.38$ & $-2.31\pm0.17$  & $1.30\pm0.23$\\
\hline
MAGIC & 2008 & $5.09\pm1.00$   & $-2.30\pm0.11$  & $5.06\pm0.66$ \\
\hline
VERITAS & 2010  & $7.82\pm0.80$  & $-2.19\pm0.07$  & $8.03\pm0.51$ \\
\hline
MAGIC & $2012-2015$ & $0.74\pm0.08$ & $-2.41\pm0.07$  & $0.71\pm0.06$\\
\hline
\hline
\end{tabular}
\end{center}
\end{table*}
   \begin{table}
\caption{Probability of a constant flux observed in the individual wavebands (see Fig.~\ref{mwl_lc}). Notes: $\bullet,\blacktriangle,\blacktriangledown$: VLBA data at 43\,GHz, $\square,\lozenge$: EVN data at 5\,GHz,  $\circ,\vartriangle$: EVN data at 1.7\,GHz, $\blacksquare, \star$: VERA data at 22 and 43\,GHz} \label{tab:mwl}
\begin{center}
\begin{tabular}{c|c|c}
\hline
\hline
Waveband        & Constant flux     & $\chi^{2}$/d.o.f.\\    
            &    probability    &    \\ 
\hline
\hline
HE         & 1.8$\times$10$^{-22}$     & 192/38        \\
\hline
X-rays (core)     & 7.5$\times$10$^{-15}$     & 72.29/4    \\ 
\hline
X-rays (HST-1)    & 6.5$\times$10$^{-10}$     & 48.87/4    \\ 
\hline
UV (core)        & 0.11                     & 7.64/4        \\ 
\hline
UV (HST-1)        & 1.4$\times$10$^{-6}$     & 32.63/4    \\
\hline
Radio                             &                        &            \\
VLBA$^{\bullet}$ (1.2 mas)            &0.67                     & 2.37/4     \\
VLBA$^{\blacktriangle}$ (Peak)            &0.44                     & 3.73/4     \\
VLBA$^{\blacktriangledown}$ (Down jet)        &0.88                     & 1.22/4     \\
EVN$^{\square}$ (Core)            &0.40                     & 5.15/5     \\
EVN$^{\circ}$ (Core)        & 1.1$\times$10$^{-8}$     & 39.95/3    \\
EVN$^{\lozenge}$ (HST-1)            & 1.5$\times$10$^{-13}$     & 69.27/5    \\
EVN$^{\vartriangle}$ (HST-1)    & 0.12                     & 5.85/5        \\
VERA$^{\blacksquare}$ (peak)     &   1.95$\times$10$^{-273}$  &  1360/23 \\
VERA$^{\star}$ (peak)     &   5.90$\times$10$^{-60}$  &  287.1/5 \\
\hline
\hline
\end{tabular}
\end{center}
\end{table}
\subsection{Multiwavelength light curves}  \label{subsec:lightcurves}
The variability of the VHE gamma-ray flux is investigated at different timescales. The mean integral flux of each year, which is obtained by a fit with a constant is reported in Table~\ref{lc1-table}. 
No variability is observed across the dataset, the only exception being a hint of variability on a daily scale observed in 2013 (the probability for a constant flux is 0.3\%).
For the other years the light curves are found to be compatible with a constant flux with a probability higher than 38\%.
Assuming an additional systematic uncertainty of 11\% of the measured flux~\citep{2016APh....72...76A} for the 2013 data, a probability for a constant flux of 0.9\% is obtained.\par
To compare the new data collected between 2012 and 2015 with previous M\,87 observations, the integral flux is calculated for $400\,\mathrm{GeV}<E<1$\,TeV and compared with previous MAGIC, H.E.S.S., and VERITAS observations \citep{2006Sci...314.1424A, 2008ApJ...679..397A,2008ApJ...685L..23A,2012A&A...544A.142A,2012ApJ...746..141A}; see Table~\ref{tab:spec}.
The integral flux level between 2012 and 2015 is compatible with those observed with H.E.S.S.\ in 2004~\citep{2006Sci...314.1424A} and with MAGIC between 2005 and 2007~\citep{2012A&A...544A.142A}, when the source was in a low emission state in the TeV band, which can be defined as some 5--10\% of the Crab Nebula flux at energies $400\,\mathrm{GeV}<E<1$\,TeV. The low emission state is clearly separated from the flaring periods in 2005, 2008 and 2010 \citep{2009Sci...325..444A,2012ApJ...746..151A,2012AIPC.1505..586B}.\par
The MWL light curve of M\,87 between 2012 and 2015 is shown in Fig.~\ref{mwl_lc}.
Both the core and the innermost knot HST-1 in the jet are found to be in low emission state. Table~\ref{tab:mwl} shows the constant flux probability and $\chi^{2}/\mathrm{d.o.f.}$ from HE to radio data. 
At lower frequencies, variability is observed for HST, {\it Chandra}, EVN 1.7\,GHz core and 5\,GHz HST-1 data, as well as VERA peak data at 22 and 44\,GHz. No clear variability is found for the EVN 1.7\,GHz HST-1 and 5\,GHz core data, VLBA core and jet data, as well as for the high-energy (HE; 100\,MeV\,$<$\,E\,$<$\,100\,GeV) gamma-ray data.\par 
The optical polarimetry data suggest a long-term rotation of the electric vector polarization angle (EVPA) from $\sim$0$^{\circ}$ to  $\sim$400$^{\circ}$, while the polarization stays in general at the rather low level of less than 5\%, except some higher polarization of up to $\sim$25\% around the beginning of the MAGIC observation period in 2012.
Since the EVPA in blazars depends on the orientation of the shocks and the magnetic field threading it, EVPA provides an important tool to understand the acceleration mechanism of the shocked plasma. In recent studies, EVPA swings larger than 180$^{\circ}$ simultaneous to gamma-ray emission have been interpreted as additional evidence for an helical structure of the magnetic field \citep{2010ApJ...710L.126M,2010Natur.463..919A,2018A&A...619A..45M}. However, in the present case, the EVPA rotation happens over several weeks (approximately from MJD~56704 to MJD~56824 -- 16 February to 16 June 2014), making it difficult to find a connection to the activity in the other bands.
EVPA rotations can be due to reconnection in the emission region of the jet during a high-activity state. However, no flare has been observed in the present datasample to be associated to a reconnection event.

\subsection{SED in the TeV band} \label{subsec:spectrum}
\begin{figure*}
  \centering
      \includegraphics[width=12cm]{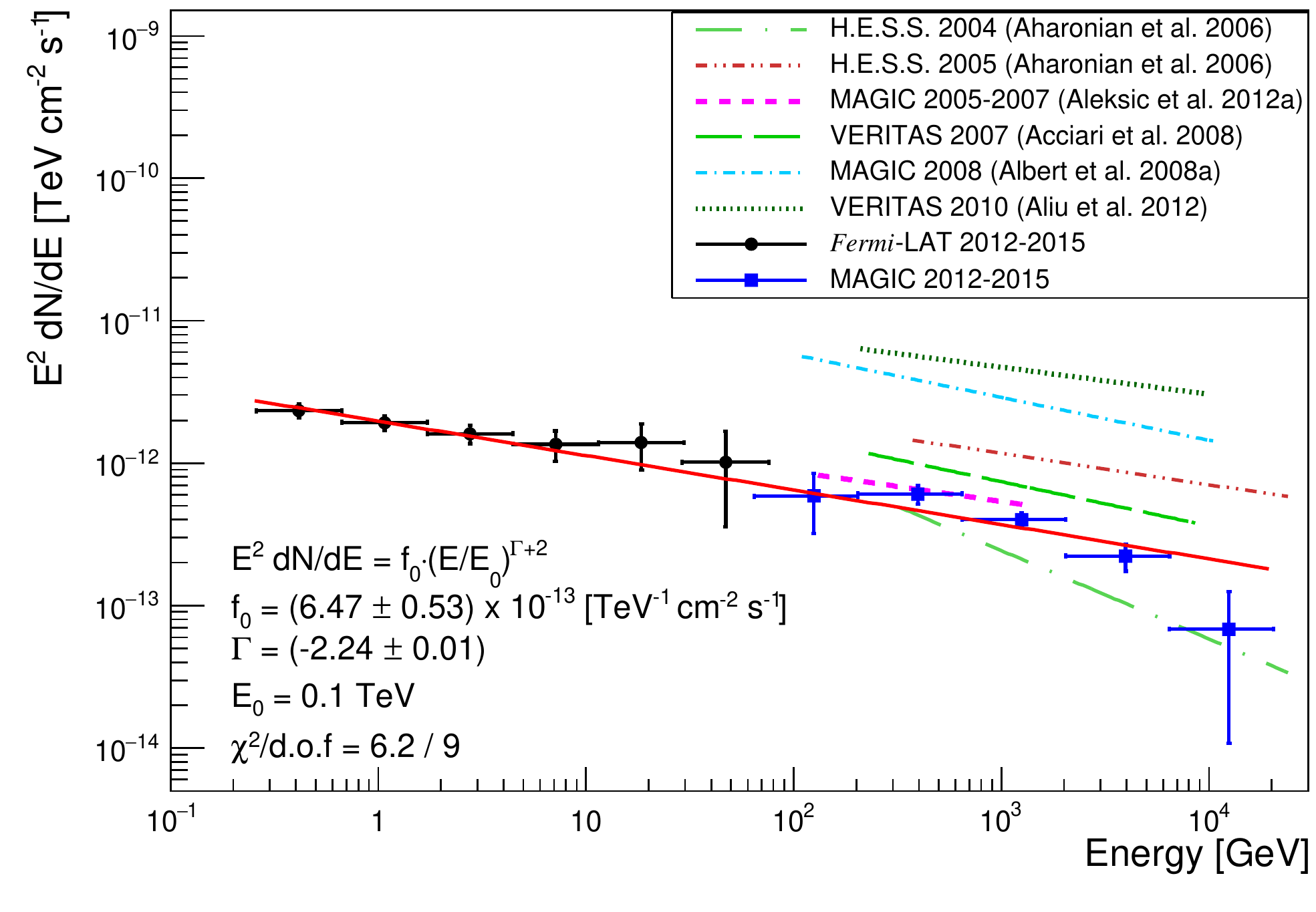}
    \caption{Combined MAGIC (blue filled squares) and quasi-simultaneous \textit{Fermi}-LAT (black filled circles) SED between 2012 and 2015, to which a simple power law (solid red line) is fitted. For comparison, the simple power-law fits describing the averaged SEDs of the low emission states observed in 2004 with H.E.S.S.~\citep[green long-dashed-dotted line;][]{2006Sci...314.1424A} and 2005-2007 with MAGIC~\citep[magenta dashed line;][]{2012A&A...544A.142A}, and of the flaring state observed in 2005 with H.E.S.S.~\citep[red short dashed-dotted line;][]{2006Sci...314.1424A}, in 2007 and 2010 with VERITAS~\citep[green long-dashed and dotted line;][]{2008ApJ...679..397A,2012ApJ...746..141A} and in 2008 with MAGIC~\citep[blue short-dashed-dotted line;][]{2008ApJ...685L..23A} are shown). 
    }%
   \label{fig:fermi-spec} 
 \end{figure*}
The VHE gamma-ray SED observed with MAGIC is well described\footnote{$\chi^{2}$/d.o.f. =
3.87/3, corresponding to a fit probability of 28\%.} by a power law of the form $E^{2} dN/dE =
f_{0 \mathrm{VHE}}\cdot(E/E_{0 \mathrm{VHE}})^{\Gamma_{\mathrm{VHE}}+2}$, with the flux normalization $f_{0 \mathrm{VHE}}$ being $(4.31 \pm 0.33)\times10^{-13}\,\mathrm{cm}^{-2}\,\mathrm{s}^{-1}\,\mathrm{TeV}^{-1}$, at a  decorrelation energy\footnote{The decorrelation energy corresponds to the energy at which the correlation between the flux normalization and spectral index is minimum.} $E_{0 VHE}$ of 784\,GeV and a spectral index $\Gamma_{VHE}$ equal to $-2.41\pm0.07$. The errors quoted here are only statistical. The observed spectrum is not significantly affected by the extragalactic background light (EBL) absorption due to the proximity of M\,87; for this source, the production of electron-positron pairs by interaction with the EBL becomes significant at higher energies, above 10~TeV~\citep{2007ApJ...671...85N}. 
The spectral indices and differential fluxes at the decorrelation energy from previous VHE gamma-ray observations are reported in Table~\ref{tab:spec}. All spectral indices observed both during high and low emission states are mostly compatible within the statistical errors. The differential flux between 2012 and 2015 is on a similar level as during the low emission states reported by MAGIC between 2005 and 2007, and by H.E.S.S. in 2004, whereas it is lower with respect to the flux level observed during the flaring states in 2005, 2008 and 2010.\par
The averaged VHE-HE gamma-ray SED between 2012 and 2015 is shown in Fig.~\ref{fig:fermi-spec}.
A simple power law (red line in Fig.~\ref{fig:fermi-spec}) is fit to the combined VHE-HE gamma-ray SED, yielding a spectral index and a flux normalization of $\Gamma=-2.24\pm0.01$ and $f_{0(E=100\,\mathrm{GeV})}=(6.47 \pm 0.53)\times10^{-13}\,\mathrm{cm}^{-2}\,\mathrm{s}^{-1}\,\mathrm{TeV}^{-1}$, respectively, at normalization energy $E_{0}$ = 100\,GeV.
While \textit{Fermi}-LAT data are collected throughout the four years, MAGIC data for reasons of visibility of the source are taken each year during the December-July time window.
\section{SED modelling}\label{sec:sed}
In \citet{2008MNRAS.385L..98T} a structured-jet model~\citep{2005A&A...432..401G} is applied, assuming a jet with a fast spine and a slower layer and thus two zones to explain the TeV flares. \citet{2012A&A...544A.142A} applies this scenario to model the low-state SED in 2005 to 2007.~\citet{2008A&A...478..111L} propose that the flaring emission would occur while the jet is collimating, and~\citet{2005ApJ...634L..33G} while it is decelerating. An alternative process to explain these VHE gamma-ray flares was proposed by~\citet{2010MNRAS.402.1649G}, which is based on misaligned mini-jets driven by magnetic reconnection moving within the jet at relativistic velocities. ~\citet{2009Ap&SS.321...57I} propose a two-step acceleration model to TeV energies involving initial particle acceleration within the accretion disk and then further centrifugal acceleration in the rotating magnetosphere. \citet{2011ApJ...730..123L} discuss the variable TeV emission possibly to be produced in a starved magnetospheric region.\par
Despite the long-term investment in M\,87 monitoring, the production site of TeV gamma rays remains unclear, with strong hints, however, that it will be close to the core of the jet (some $40-100\,R_\mathrm{S}$ from the black hole), suggested by the correlation with radio and X-ray activities \citep{2008ApJ...679..397A}. HST-1 has also been discussed as a possible production site of the TeV emission because of a rapid TeV flare in 2006 \citep{2006ApJ...640..211H} that coincides with an enhanced X-ray flux of  this knot, while no enhanced flux from the nucleus is observed. However, the association with HST-1 seems unlikely given the absence of radio or X-ray short-timescale variabilities detected from this region, whereas several occasions of gamma-ray variability on daily timescales were seen. Another caveat to this interpretation is given by the fact that VLBA data have shown the compact knots in HST-1 to be essentially unresolved  approaching the size limits set by the TeV-emission variability, as reported in \cite{2007ApJ...663L..65C}.\par
\begin{figure*}
\centering
\includegraphics[width=14cm]{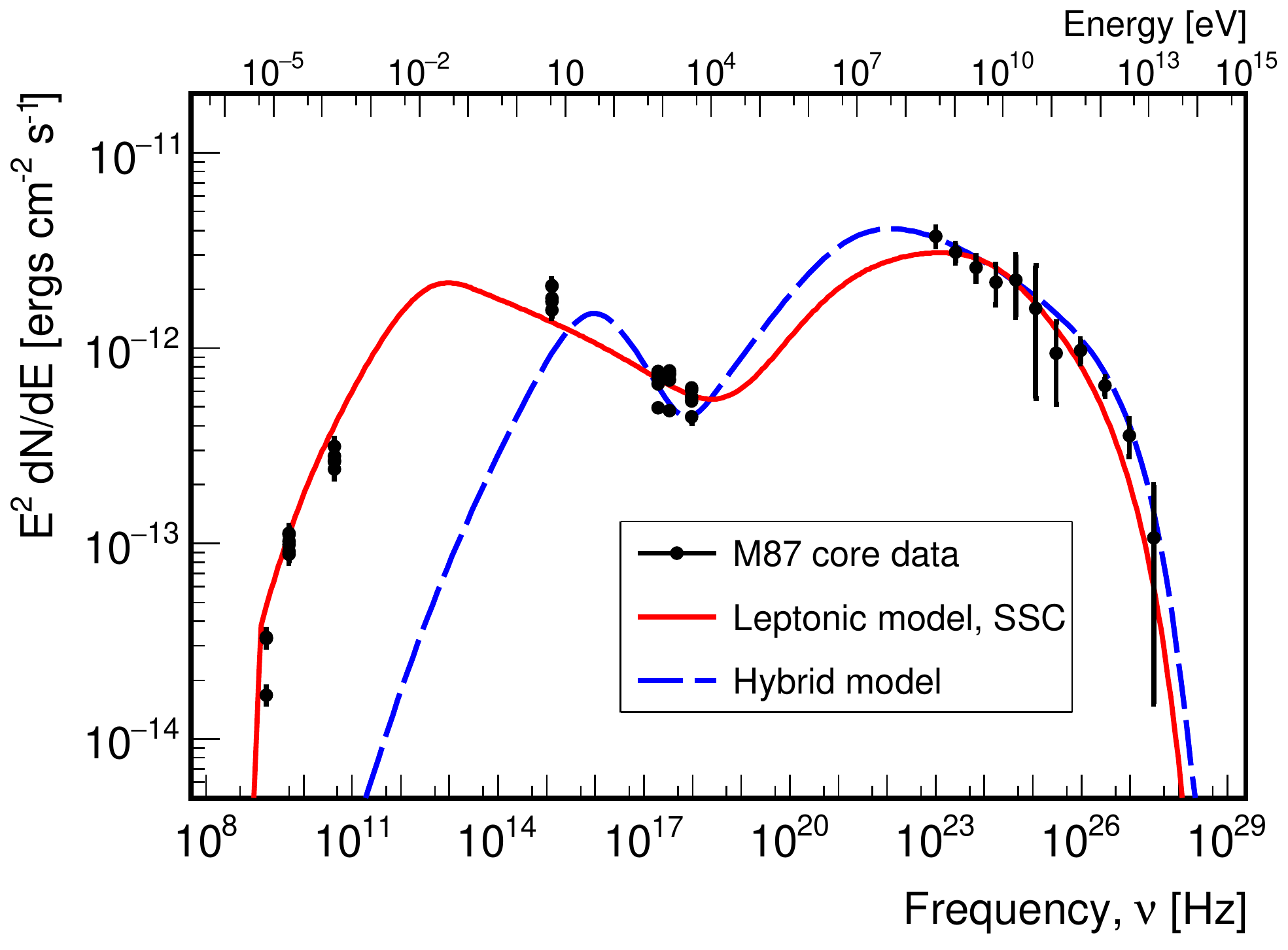}
\caption{MWL SED of the radio core of M\,87 compiled from quasi-simultaneous 2012--2015 observations (black points). VHE gamma-ray observations by MAGIC are combined with HE gamma-ray data from \textit{Fermi}-LAT, X-ray data from \textit{Chandra}, NUV data from HST, radio data at 1.7\,GHz and 5\,GHz provided by the EVN and at 43\,GHz by VLBA. The models represent two possible scenarios: in the leptonic scenario (red solid line) the high-energy component is dominated by the SSC emission, whereas in the hybrid scenario (blue dashed line) the high energy emission is dominated  by the synchrotron radiation of relativistic protons.}
\label{fig:SED} 
\end{figure*}

In this paper, the question whether a self-consistent modelling of a single emission region can explain the observed data in the low flux state and what consequences such scenario would have is investigated.
Fig.~\ref{fig:SED} shows the MWL SED of the radio core of M\,87. To build the SED, quasi-simultaneous data are used, taken between 2012 to 2015 from MAGIC, \textit{Fermi}-LAT, Chandra, HST, EVN at 1.7~GHz and 5~GHz, and VLBA at 43~GHz. NUV data are corrected for interstellar extinction following ~\citet{1989ApJ...345..245C}. No averaging is performed for the low-energy data from radio to X-ray, where variability has been observed. \par
As no clear variability in the TeV regime is detected, the model constraints are relaxed with respect to those for flaring states. The acceleration and emission zones are assumed to be directly connected representing the downstream region of a shock front.
Additionally, the relevant parts in the quasi-simultaneous MWL data connected to the TeV gamma-ray emitting zone are assumed to be the HE gamma-ray data from \textit{Fermi}-LAT and the X-ray data detected for the core region by \textit{Chandra}. In general, radio emission tends to have a spatially-extended emission region, and can be subject to synchrotron self-absorption in the core region due to high magnetic field strength leading to high-opacity in this band. However, usually this absorption effect is more severe for hadronic models than it is for leptonic scenarios, where a less strong magnetic field is required. We therefore assume the radio emission observed by VLBA and EVN originate from a larger region. The NUV data from HST detected for the core of M\,87 presumably have origin from a region much closer to the black hole and where the jet is launched, as suggested by General Relativistic Radiation Magnetohydrodynamics~\citep[see, e.g.,][]{2015ApJ...807...31R}.
Therefore, we concentrate on the X-ray, GeV and TeV data rather than the radio-to-optical data for our modelling.\par
First, the leptonic model is applied to account for the broadband spectrum of M\,87 with the numerical code in \citealt{2014ApJ...780...64A}, neglecting Fermi-II acceleration~\citep[see also,][]{2015ApJ...808L..18A,2018ApJ...861...31A}. 
The code calculates the temporal evolution of the electron and photon energy distributions in the plasma rest frame along the jet (at radius $R$ from the black hole), which is similar to the {\tt BLAZAR} code in \citealt{2003A&A...406..855M}
\citep[for application examples see e.g.,][]{2008ApJ...672..787K,2012ApJ...754..114H}.
Here, a steady conical outflow is assumed, in which the temporal evolution along the jet is equivalent to the radial evolution. The conically expanding jet naturally leads to adiabatic cooling of electrons, which is a similar effect to the electron escape in one-zone steady models. In this 1-D code, the parameter for the electron escape is not required. 
The magnetic field decreases as $B=B_0 (R/R_0)^{-1}$.
The macroscopic model parameters are the Lorentz factor $\Gamma_{L}$, the initial radius $R_0$ (distance from the black hole), the initial magnetic field $B_0$, the electron luminosity $L_{\rm e}$ (including the counter jet), the jet opening angle $\theta_{\rm j}$, and the viewing angle $\theta_{\rm v}$.
Here, $\Gamma_{L}=3$, $\theta_{\rm j}=1/\Gamma_{L}=19^\circ$ and $\theta_{\rm v}=15^\circ$ half-opening angle of the jet are adopted~\citep{1999ApJ...520..621B,2009Sci...325..444A,2018ApJ...855..128W}, with the half-opening angle well below the average apparent full opening angle inferred from radio observations~\citep{2018ApJ...855..128W}.
Electrons are injected during the dynamical timescale $R_0/(c \Gamma_{L})$ in the plasma rest frame. In this timescale, the injection rate into a given volume $V \propto R^2$ is constant. The evolutions of the electron energy distribution and photon emission are calculated as far as $R=30 R_0$, taking into account synchrotron emission and inverse Compton scattering with the Klein--Nishina effect, gamma-gamma absorption, secondary pair injection, synchrotron self-absorption, and adiabatic cooling.
The model parameters for the electron injection spectrum are: the minimum and maximum electron Lorentz factors ($\gamma_{\rm min}$ and $\gamma_{\rm max}$), the location of the break in the electron energy distribution ($\gamma_{\rm br}$), and power-law indices $p_1$ and $p_2$ for below and above $\gamma_{\rm br}$ respectively.
The parameter values are summarized in Table \ref{tab:fit_parameters_together}.\par
Second, a hybrid model is applied assuming that protons and electrons are accelerated in the jet.
The fully time-dependent implementation is based on the geometry of \citet{2010IJMPD..19..887W}. The acceleration mechanism and the implementation of all leptonic processes are adopted from \citet{2016ApJ...829...56R} and the photo-hadronic framework is implemented following \citet{2010ApJ...721..630H}. The acceleration of particles is closely modeled to the Fermi-I acceleration. Under the assumption that the particle distribution is quickly reaching isotropy in the downstream region of the shock, the model follows the evolution of the injected, monoenergetic particle distribution towards a power law. The shape of the particle distribution and the relevant timescales follow consistently from the input shock parameters.
\begin{table*} 
\caption{Parameters used for the models shown in Fig.~\ref{fig:SED}. Parameters of the Leptonic model are described in the text. Parameters of the Hybrid model are explained in Table~\ref{tab:parameters}.}
\label{tab:fit_parameters_together} 
\begin{center}
\begin{tabular}{c|c|c|c|c|c|c|c|c|c|c|c} 
\hline\hline
\multirow{6}{*}{Leptonic} & & & & & & & & & & & \\
 & $\Gamma_{L}$ & $R_{0}$ & $B_{0}$ & 
$L_{\rm e}$ & $\theta_{\rm j}$ & $\theta_{\rm v}$
& $\gamma_{\rm min}$ & $\gamma_{\rm br}$ & $\gamma_{\rm max}$ & $p_{1}$
& $p_{2}$\\ 
& & $[10^{17} \mathrm{cm}]$ & $[\mathrm{mG}]$ & $[10^{44} \mathrm {erg~s}^{-1}]$  & $[^{\circ}]$ & $[^{\circ}]$&
   & & & \\
\cline{2-12}
& & & & & & & & & & & \\
& $3$ & $4.0$ & $3.1$ & $2.4$ & $19$ & $15$ & $500$ & $1.4\times10^{4}$ & $3.0\times10^{7}$ & $1.9$ & 3.2 \\
& & & & & & & & & & & \\
 \hline\hline
\multirow{6}{*}{Hybrid} & & & & & & & & & & & \\
 & $R_{\mathrm{acc}} $ & $R_{\mathrm{rad}}$ & $N_{\mathrm{inj}}^{\mathrm{p}}$ & 
$\gamma_{\mathrm{inj}}^{\mathrm{p}}$ & $N_{\mathrm{inj}}^{\mathrm{e}}$ & $\gamma_{\mathrm{inj}}^{\mathrm{e}}$
& $B$ & $r$ & $\eta_{e}$ & $\mathcal{D}$ & \\ 
 & $[10^{12}\mathrm{cm}]$ & $[10^{15}\mathrm{cm}]$ & $[10^{42}\mathrm{cm s^{-1}}]$ & &
$[10^{40}\mathrm{cms^{-1}}$] & $[10^{3}]$ & $[G]$ & & & &\\ 
& & & & & & & & & & & \\
\cline{2-12}
& & & & & & & & & & & \\
& $35$ & $30$ & $8{.}77$ & $10$ & $2$ & $10$ & $3$ & $3{.}50$ & $1$ & $5.3$ & \\ 
& & & & & & & & & & & \\
\hline\hline 
\end{tabular} 
\end{center}
\end{table*}
\begin{table}
\caption{Description of the parameters used in the hybrid modelling of the broadband SED}
\label{tab:parameters}
\begin{center}
\begin{tabular}{cl} 
\hline\hline 
\rule{0pt}{3ex} 
$R_{\mathrm{rad}}$ & size of the radiation zone\\ $R_{\mathrm{acc}}$ & size of the acceleration zone\\
$\gamma_{\mathrm{inj}}^{\mathrm{species}}$ & the energy injected
\\ 
$B$ & magnetic field strength\\ 
$\eta$ & particle diffusion coefficient\\ 
$V_S$ & shock speed\\ 
$r$ & shock compression ratio\\ 
$\mathcal{D}$ & Doppler factor\\
\hline\hline 
\end{tabular} 
\end{center}
\end{table} 
The simulated SED, computed with the hybrid model that best describes the observed broadband SED, is shown in Fig.~\ref{fig:SED} (blue dashed line) together with the available quasi-simultaneous data.
It is not clear whether a unique set of parameters exists for describing the SED and the high dimensionality of the parameter space does not allow for $\chi^{2}$ fitting. The community standard is therefore to optimize the SED modelling by manual parameter changes until data points and especially slopes are agreeing with the observed SED. 
In the hybrid model, the radio-to-X-ray radiations originate from synchrotron emission of electrons. The emission at higher energies, due to the high magnetic field and the assumption of protons being injected into the acceleration zone, are dominated by synchrotron emission of protons in this case.
The parameters used for the presented fit are summarized in Table~\ref{tab:fit_parameters_together}, and the most important parameters of the model are described in Table \ref{tab:parameters}.  
The shock speed is set to $V_S= 0.1c$ and Fermi-II acceleration is neglected. The spatial diffusion
coefficient $\eta$ is mass dependent, we will use \mbox{$\eta_{e} = \eta$} as
reference value and calculate the proton coefficient as \mbox{$\eta_{p} =m_{p}/m_{e}
\eta$}. The escape time from each region is calculated as $t_{\mathrm{esc}}=\eta R/c$ and the acceleration timescale follows, depending on the spectral index, from $t_{\mathrm{acc}} =\eta R_{\mathrm{acc}}/c \gtrsim t_{\mathrm{esc}}$
\citep[for further reading see, e.g.,][]{2004PASA...21....1P}.
It has to be noted that \textit{Chandra} data are in the sensitive area of the transition between the two bumps ($10^{17} - 10^{18}$\,Hz) in the hybrid model. 
In addition, the number of parameters is higher than in the leptonic model and the two components (self-Compton from electrons and synchrotron from protons) are basically independent of each other, as their densities are assumed to be uncorrelated. The hybrid model covers electrons and
protons simultaneously. Hence self Compton and proton synchrotron emission exist within the same model.\par
The MAGIC observations reveal that the GeV--TeV emission is compatible with a single emission component, either self-Compton from the leptons or synchrotron from protons.
The component from radio to X-rays is also explained by the synchrotron component of leptons in both scenarios.
However, in the leptonic scenario, radio and X-rays would originate from the same region as the GeV--TeV data whereas in the hybrid scenario they would come from a different and a much larger region.
The required electron luminosity ($L_{\rm e}=2.4 \times 10^{44}~\mbox{erg}~\mbox{s}^{-1}$
including the counter jet) is comparable to the total jet power
estimated from the large scale radio structure \citep{2000ApJ...543..611O}.
However, the relatively high SSC flux requires a very low magnetization: the energy-density ratio of the magnetic field to the non-thermal electrons is $5.2 \times 10^{-5}$ at $R=2 R_0$, which ismuch smaller than the values found in other blazars~\citep[$10^{-1}-10^{-2}$,][]{2018ApJ...861...31A}.

\citet{2012ApJ...745L..28A} claim that the radio image of the M\,87 jet is consistent with a parabolic streamline, also confirmed by \citet{2013ApJ...775...70H}, which supports the magnetically-driven jet model \citep{2009MNRAS.394.1182K}.
Later, \citet{2014ApJ...781L...2A} show that the gradual acceleration through a distance of $10^6$ times the Schwarzschild radius also supports the magnetically driven jet model.
In addition \citet{2015ApJ...803...30K} point out that the radio data at 230\,GHz obtained by the Event Horizon Telescope implies a magnetically dominated jet.
Those results seem inconsistent with the very low magnetization at $\sim$200 times the Schwarzschild radius indicated by the broadband SED.
Other than the large-scale component constrained by radio observations, a very low-magnetized emission region is required to explain the gamma-ray spectrum by the leptonic model.
Such a very low jet magnetization in the emitting region (the energy in leptons is 5 orders of magnitude higher than the energy in the magnetic field) is very far from the equipartition scenario. This suggests an emission region far from the core of the jet or a very efficient mechanism to convert Poynting flux into a matter dominated jet. \par
A similar low magnetization in leptonic modelling of blazars is also reported in \cite{2016MNRAS.456.2374T}, where the sample of BL Lacs used for the study were characterised by a small value of the magnetic energy density and a relatively large Doppler factor, not easy to be explained in a one-zone SSC scenario, but justified in a spine-layer model, supporting the hypothesis of structured jets in BL Lac objects.\par 
New results from the EHT \citep{2019ApJ...875L...5E} favor the hypothesis of the accretion disk as the origin of the observed emission at 230\,GHz. However, given the fact that many of the used models produce images consistent with the EHT data, they suggests that the image shape is mainly controlled by gravitational lensing and the spacetime geometry, rather than details of the plasma physics.\par
The physical picture of the M\,87 jet remains unrevealed yet, which provides great motivation of future observational and theoretical study for this object and blazars in general.

\section{Conclusions}\label{sec:conclusions}

MAGIC has monitored M\,87 from 2012 to 2015 for a total of 156~hrs after quality cuts. The source is detected in a low state every yearly campaign between 2012 and 2015. No clear variability is observed in the 2012, 2014, and 2015 data on daily and $\sim$20-days timescales. A hint of variability ($\sim$3$\,\sigma$ level) is found in 2013 data on a daily timescale, remaining at a similar significance level even when variable systematic uncertainties of the MAGIC measurements are taken into account. The VHE gamma-ray flux level above 300\,GeV between 2012 and 2015 is the lowest flux observed since 2005.\par
No clear variability was found at lower frequencies for the VLBA core and the jet data, nor for the EVN 1.7-GHz and 5-GHz data for HST-1 and the core respectively. However, variability was observed in HST, {\it Chandra}, EVN 1.7-GHz core and EVN 5-GHz HST-1 data, and VERA peak data at 22 and 44\,GHz.
The optical polarimetry data suggest a long-term rotation from $\sim0^{\circ}$ to  $\sim400^{\circ}$. The polarization stays in general at a rather low level, below 5\%, except some higher polarization of up to $\sim25\%$ around the beginning of the MAGIC observation period in 2012.\par
The energy spectrum of M\,87 is found to have the same shape (within the statistical uncertainties) during the observations presented here and during the TeV flares observed in the past. Remarkably, the combination with \textit{Fermi}-LAT data in GeV energies reveals a continuous photon spectrum over 5 orders of magnitude in energy, which is consistent with a simple power law.\par
The broadband SED is found to be consistent with
leptonic and hybrid emission scenarios where the GeV--TeV component corresponds to self-Compton or synchrotron radiation from leptons and hadrons, respectively.
An important result from the leptonic modelling is that the required jet magnetization in the emitting region is very low (the energy in leptons is 5 orders of magnitude higher than that in the magnetic field) and thus very far from the equipartition scenario. This either implies an emission region far from the core of the jet or a very efficient mechanism to convert Poynting flux into a matter dominated jet. \par
Both leptonic and hybrid model provide a good description of the data. However, the hybrid scenario is more consistent with the HE and VHE gamma-ray part of the SED. For this reason, it is preferred in describing the present low state of the source..\par
Further dense and precision MWL observations of M\,87 are necessary to unveil the nature of the emission and its spatial location.

\section*{Acknowledgements}
%
%
We are grateful to Dr. Filippo D'Ammando and Dr. C.~C. Cheung for valuable discussions and their useful comments and suggestions. \\
We would like to thank the Instituto de Astrof\'{\i}sica de Canarias for the excellent working conditions at the Observatorio del Roque de los Muchachos in La Palma. The financial support of the German BMBF and MPG, the Italian INFN and INAF, the Swiss National Fund SNF, the ERDF under the Spanish MINECO (FPA2015-69818-P, FPA2012-36668, FPA2015-68378-P, FPA2015-69210-C6-2-R, FPA2015-69210-C6-4-R, FPA2015-69210-C6-6-R, AYA2015-71042-P, AYA2016-76012-C3-1-P, ESP2015-71662-C2-2-P, FPA2017-90566-REDC), the Indian Department of Atomic Energy, the Japanese JSPS and MEXT and the Bulgarian Ministry of Education and Science, National RI Roadmap Project DO1-153/28.08.2018 is gratefully acknowledged. This work was also supported by the Spanish Centro de Excelencia ``Severo Ochoa'' SEV-2016-0588 and SEV-2015-0548, and Unidad de Excelencia ``Mar\'{\i}a de Maeztu'' MDM-2014-0369, by the Croatian Science Foundation (HrZZ) Project IP-2016-06-9782 and the University of Rijeka Project 13.12.1.3.02, by the DFG Collaborative Research Centers SFB823/C4 and SFB876/C3, the Polish National Research Centre grant UMO-2016/22/M/ST9/00382 and by the Brazilian MCTIC, CNPq and FAPERJ. \\
The Very Long Baseline Array is operated by the National Radio Astronomy Observatory, which is a facility of the National Science Foundation operated under cooperative agreement by Associated
Universities, Inc.\\
The work of F.Massaro is supported by the ``Departments of Excellence 2018 - 2022'' Grant awarded by the Italian Ministry of Education, University and Research (MIUR) (L. 232/2016). His research has made also use of resources provided by the Compagnia di San Paolo for the grant awarded on the BLENV project (S1618\_L1\_MASF\_01) and by the Ministry of Education, Universities and Research for the grant MASF\_FFABR\_17\_01. This investigation is supported by the National Aeronautics and Space Administration (NASA) grants GO4-15096X, AR6-17012X and GO6-17081X. \\
This work made use of the Swinburne University of Technology software
correlator, developed as part of the Australian Major National Research
Facilities Programme and operated under licence. \\
Based in part on observations made with the NASA/ESA Hubble Space Telescope, obtained at the Space Telescope Science Institute, which is operated by the Association of Universities for Research in Astronomy, Inc., under NASA contract NAS5-26555. These observations are associated with programmes GO 12293, 12671, 13061, and 13759.\\
Dan Harris passed away on December 6th, 2015.
His career spanned much of the history of radio and X-ray astronomy.
His passion, insight, and contributions will always be remembered.


\section*{Affiliations}
\noindent
{\it
$^{1}$ {Inst. de Astrof\'isica de Canarias, E-38200 La Laguna, and Universidad de La Laguna, Dpto. Astrof\'isica, E-38206 La Laguna, Tenerife, Spain}\\
$^{2}$ {Universit\`a di Udine, and INFN Trieste, I-33100 Udine, Italy}\\
$^{3}$ {National Institute for Astrophysics (INAF), I-00136 Rome, Italy}\\
$^{4}$ {ETH Zurich, CH-8093 Zurich, Switzerland}\\
$^{5}$ {Technische Universit\"at Dortmund, D-44221 Dortmund, Germany}\\
$^{6}$ {Croatian Consortium: University of Rijeka, Department of Physics, 51000 Rijeka; University of Split - FESB, 21000 Split; University of Zagreb - FER, 10000 Zagreb; University of Osijek, 31000 Osijek; Rudjer Boskovic Institute, 10000 Zagreb, Croatia}\\
$^{7}$ {Saha Institute of Nuclear Physics, HBNI, 1/AF Bidhannagar, Salt Lake, Sector-1, Kolkata 700064, India}\\
$^{8}$ {Centro Brasileiro de Pesquisas F\'isicas (CBPF), 22290-180 URCA, Rio de Janeiro (RJ), Brasil}\\
$^{9}$ {IPARCOS Institute and EMFTEL Department, Universidad Complutense de Madrid, E-28040 Madrid, Spain}\\
$^{10}$ {University of \L\'od\'z, Department of Astrophysics, PL-90236 \L\'od\'z, Poland}\\
$^{11}$ {Universit\`a  di Siena and INFN Pisa, I-53100 Siena, Italy}\\
$^{12}$ {Deutsches Elektronen-Synchrotron (DESY), D-15738 Zeuthen, Germany}\\
$^{13}$ {Istituto Nazionale Fisica Nucleare (INFN), 00044 Frascati (Roma) Italy}\\
$^{14}$ {Max-Planck-Institut f\"ur Physik, D-80805 M\"unchen, Germany}\\
$^{15}$ {Institut de F\'isica d'Altes Energies (IFAE), The Barcelona Institute of Science and Technology (BIST), E-08193 Bellaterra (Barcelona), Spain}\\
$^{16}$ {Universit\`a di Padova and INFN, I-35131 Padova, Italy}\\
$^{17}$ {Universit\`a di Pisa, and INFN Pisa, I-56126 Pisa, Italy}\\
$^{18}$ {ICRANet-Armenia at NAS RA, 0019 Yerevan, Armenia}\\
$^{19}$ {Centro de Investigaciones Energ\'eticas, Medioambientales y Tecnol\'ogicas, E-28040 Madrid, Spain}\\
$^{20}$ {Universit\"at W\"urzburg, D-97074 W\"urzburg, Germany}\\
$^{21}$ {Finnish MAGIC Consortium: Finnish Centre of Astronomy with ESO (FINCA), University of Turku, FI-20014 Turku, Finland; Astronomy Research Unit, University of Oulu, FI-90014 Oulu, Finland}\\
$^{22}$ {Departament de F\'isica, and CERES-IEEC, Universitat Aut\`onoma de Barcelona, E-08193 Bellaterra, Spain}\\
$^{23}$ {Japanese MAGIC Consortium: ICRR, The University of Tokyo, 277-8582 Chiba, Japan; Department of Physics, Kyoto University, 606-8502 Kyoto, Japan; Tokai University, 259-1292 Kanagawa, Japan; RIKEN, 351-0198 Saitama, Japan}\\
$^{24}$ {Inst. for Nucl. Research and Nucl. Energy, Bulgarian Academy of Sciences, BG-1784 Sofia, Bulgaria}\\
$^{25}$ {Universitat de Barcelona, ICCUB, IEEC-UB, E-08028 Barcelona, Spain}\\
$^{26}$ {also at Port d'Informaci\'o Cient\'ifica (PIC) E-08193 Bellaterra (Barcelona) Spain}\\
$^{27}$ {also at Dipartimento di Fisica, Universit\`a di Trieste, I-34127 Trieste, Italy}\\
$^{28}$ {also at INAF-Trieste and Dept. of Physics \& Astronomy, University of Bologna}\\
$^{29}$ {Centre for Space Research, North-West University, Potchefstroom 2520, South Africa}\\ 
$^{30}$ {INAF Osservatorio Astronomico di Padova, vicolo dell' Osservatorio 5, 35122, Padova, Italy}\\
$^{31}$ {ICRR, University of Tokyo, Kashiwa-shi, Chiba 277-8582, Japan}\\ 
$^{32}$ {Mizusawa VLBI Observatory, National Astronomical Observatory of Japan, 2-12 Hoshigaoka-cho, Mizusawa, Oshu, Iwate 023-0861, Japan}\\
$^{33}$ {Department of Astronomical Science, The Graduate University for
Advanced Studies (SOKENDAI), 2-21-1 Osawa, Mitaka, Tokyo 181-8588,
Japan}\\
$^{34}$ {Smithsonian Astrophysical Observatory, 60 Garden Street, Cambridge, MA 02138, USA}\\
$^{35}$ {INAF Istituto di Radioastronomia, via Gobetti 101, 40129 Bologna, Italy}\\
$^{36}$ {Astrophysics Research Institute, Liverpool John Moores University, Liverpool, UK}\\ 
$^{37}$ {University of Texas Rio Grande Valley, Department of Physics and Astronomy, One West University Blvd. Brownsville, TX 78520, USA}\\ 
$^{38}$ {Dipartimento di Fisica, Universit\`a degli Studi di Torino, via Pietro Giuria 1, I-10125 Torino, Italy}\\  
$^{39}$ {INFN, Sezione di Torino, I- 10125 Torino, Italy}\\  
$^{40}$ {INAF-Osservatorio Astrofisico di Torino, via Osservatorio 20, I-10025 Pino Torinese, Italy}\\  
$^{41}$ {Consorzio Interuniversitario per la Fisica Spaziale (CIFS), via Pietro Giuria 1, I-10125, Torino, Italy}\\ 
$^{42}$ {National Radio Astronomy Observatory, Socorro, NM 87801 USA}\\ 
$^{43}$ {Institut f\"ur Theoretische Astrophysik, Universit\"t Heidelberg}
}



\bibliographystyle{mnras}
\bibliography{references} 





\bsp	
\label{lastpage}
\end{document}